\documentstyle[12pt,doublespace]{article}
\newcommand{\Oad}{O$_{ad}\;$}
\newcommand{\Had}{H$_{ad}\;$}
\input{psfig}
\begin{document}
\title{Dynamics of Adsorbate Islands \\
       with Nanoscale Resolution}
\author{Norbert Kruse, Christian Voss, Valentin Medvedev\thanks{Present
address: Chemical Engineering, University of Washington, Seattle, 
WA 98195-1750} \\ 
{\em Chemical Physics at Surfaces and Heterogeneous Catalysis CP 243} \\
{\em Universit\'e  Libre de Bruxelles, 1050--Bruxelles, Belgium} \\
\vspace*{.1in}
Christopher Bodenstein, David Hanon, and 
Jean Pierre Boon\thanks{E-mail address: {\tt jpboon@ulb.ac.be}}\\
{\em Center for Nonlinear Phenomena and Complex Systems CP 231} \\
{\em Universit\'{e} Libre de Bruxelles, 1050--Bruxelles, Belgium}}
\date{(30 June 2000)}
\maketitle
 
\newpage

\begin{abstract}

Surface catalytic processes produce, under certain conditions, small
clusters of adsorbed atoms or groups, called {\em islands} which,
after they have been formed, move as individual entities. Here we
consider the catalytic reduction of NO with hydrogen on platinum. 
(i) Using video field ion microscopy, we observe the dynamic motion of 
small hydroxyl islands on the Pt(001) plane; despite changes in their 
morphology, the islands dimensions are confined to values corresponding 
to 10 to 30 Pt atoms suggesting cooperative effects to be in operation. 
(ii) We construct an automaton (or lattice Monte-Carlo) model on the basis 
of a set of elementary processes governing the microscopic dynamics. 
The agreement  between the simulation results and the experimental 
observations suggests a possible mechanism for the formation and dynamics 
of hydroxyl islands.

\vspace*{.1in}


\end{abstract}


\newpage

Modern surface science aims at elucidating the microscopic mechanisms of 
dynamic reaction phenomena ~\cite{wintter}. Scanning tunneling microscopy
(STM) has largely contributed to the development of the field by providing 
information at the atomic scale. For example, a recent study of the oxygen 
dissociation on the plane (111) of a platinum surface~\cite{zambel} has made 
visible the formation of adsorbate clusters called {\em islands}. 
Field ion microscopy (FIM) is an alternative method where samples are
given in the form of small three-dimensional tips exposing a number of
crystallographically different planes. The capability of the method to
image dynamic reactive phenomena at nanoscale has been 
explored~\cite{gorod,vosskr,sieben}. For example, reaction-diffusion fronts 
moving across the tip surface have been made visible thus providing 
information on the communication behavior of different planes. As compared to 
STM, the time resolution of the FIM method is only limited by the frequency 
of the video system (20~msec. here).

While the formation and dynamical behavior of islands appear as a sequence of 
processes which, in general, are difficult to explain on the basis of a 
phenomenological description, this type of system is well suited for modeling 
with Cellular Automata~\cite{boon}, which are constructed with simple rules, 
and may thereby suggest basic mechanisms for complex reactive phenomena.
Here we report on FIM experiments showing the formation of small clusters 
of hydroxyl species from a co-adsorbed \Oad/\Had layer on the (001) plane of 
a Pt tip surface during the reaction of nitric oxide (NO) with hydrogen gas, 
and, in order to gain insight into the basic mechanisms governing the fairly
complex processes producing islands, we construct an automaton (or lattice 
Monte-Carlo model) with simple rules which offers a microscopic approach 
applicable to the experimental system studied by FIM.

The experimental studies were performed with a [001]-oriented Pt tip (metal
purity 99.99~{\%}) prepared by electrochemical methods. The sample was mounted
in an all-metal ultra-high vacuum (UHV) field ion microscope (base pressure
$10^{-8}$ Pa) and cleaned using standard methods \cite{vosskr,gauss}.
Gases were used in either commercially available purity (NO: 99.5~{\%}) or
after further purification using adsorption methods (H$_2$ and Ne: better 
than 99.999~{\%} each). Ne-field ion imaging prior to reaction experiments 
indicated a clean defect-free Pt tip. As described in~\cite{vosskr},
the nearly hemispherical shape of the tip specimen is transformed into a 
pyramid during adsorption of NO gas or reaction with NO/H$_2$ gas mixtures, 
respectively. An essential feature of the pyramidal form is the appearance 
of large (111) oriented slopes and a (001) truncated top considerably 
increased in size as compared to the original state. Figure~1 shows a 
respective Ne-field ion image at 57 K obtained after the reaction studies. 
Only a few net planes are observed in the region between the (001) pole and 
the peripheral (111) planes. Subsequent to NO/H$_2$ reaction studies, 
a chemical surface analysis was performed using atom-probe techniques. 
The respective mass spectra~\cite{vossab} demonstrated the Pt tip surface to 
be oxidized, i.e. O$_{ad}$-covered. While in previous research, oscillating 
reaction phenomena with water (and nitrogen) product formation were 
examined in detail~\cite{vosskr,sieben}, we focus here on the occurrence of 
O\Had islands from an \Oad/\Had co-adsorbed layer, with particular emphasis 
on the dynamics of island formation and displacement on the Pt (001) pole. 

\begin{figure}
\centerline{
            \psfig{figure=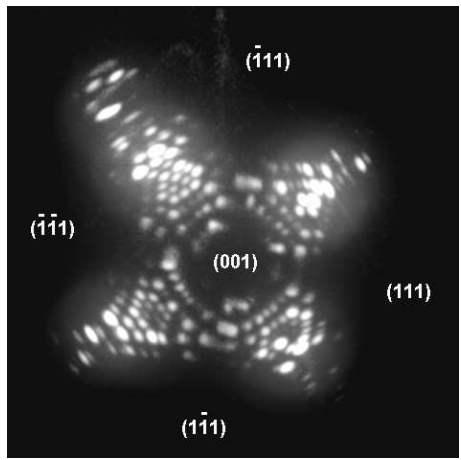,height=6cm}
           }
\vspace{4mm}
\caption{Field ion micrograph (image gas: H$_2$) of a Pt tip demonstrating a 
truncated pyramidal form with a [001] oriented top; the morphology is the 
result of a NO-induced reconstruction of an originally hemispherical Pt 
specimen; the image was obtained after field evaporation of a few surface
layers subsequent to reaction with NO/H$_2$ gas.}
\label{fig.1}
\end{figure}

The studies were performed under truly {\em in-situ}
conditions, i.e. video field ion imaging during the ongoing surface
reaction. Figure 2 shows a typical sequence of images as obtained while
exposing an initially clean, nearly hemispherical Pt tip specimen to
reactant pressures of p$_{H_2}$=4$\times$10$^{-3}$ Pa and 
p$_{NO}$=3$\times$10$^{-3}$ Pa at 500 K and 8.7 V/nm. 
After reaction-induced shape transformation into a pyramidal
morphology small islands (appearing as dark patches on a bright background
in Fig.2) with an equivalent size of 10 to 30 Pt surface atoms become
discernible on the (001) pole. Islands exclusively form at the layer edge
and move into the (001) terrace region. Up to four islands can be seen at
the same time. Mean lifetimes of several minutes are observed for
individual islands before their annihilation at the layer ledge.

\begin{figure}
\centerline{
            \psfig{figure=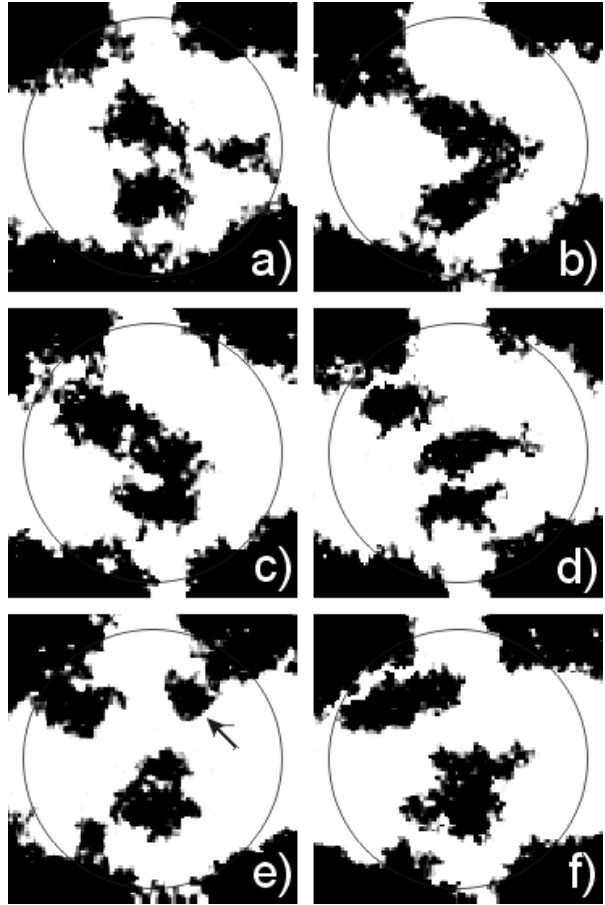,height=12cm}
           }
\vspace{4mm}
\caption{Typical sequence of video field ion images demonstrating OH$_{ad}$ 
island formation during the catalytic reduction of NO by H$_2$. 
Experimental conditions: T =  500 K,
p$_{H_2}=4\times10^{-3}$ Pa, p$_{NO}=3\times10^{-3}$ Pa, F= 8.5 V/nm.
Micrographs are treated by software allowing to optimize contrast and 
brightness; circles are understood as a guide to the eye to approximate the 
location of the (001) plane.
Snapshots taken over a total time interval of order of one sec. demonstrate 
islands displacement, collision and separation; the arrow  points 
to an island that will dissolve at the layer edge while proceeding to (f).}
\label{fig.2_new}
\end{figure}

One of the interesting features in the genesis of small islands is their
critical size. Accordingly, clusters extending over less than $\sim$ 10 atoms 
of the Pt(001) plane are not observed. Conversely, spreading over more than
$\sim$ 30 Pt atoms does not occur either. Moving clusters yet suffer some 
changes in their morphology. Their simultaneous presence on the (001) plane 
may cause collisions, however, no mergence is seen.
Under the experimental conditions applied, both NO and H$_2$ undergo
dissociation on the Pt tip surface. The dissociation probability is
strongly dependent on the local surface crystallography. Accordingly,
starting with a clean surface, the Pt (001) plane is readily covered by \Oad
(nitrogen atoms formed concomitantly during NO decomposition undergo
recombinative thermal desorption at the reaction temperature, 
T=500 K \cite{schwa,lombar,zemly}), while the rough surface planes 
accommodate both \Oad and \Had. In fact, hydrogen activation at the (001) 
layer edge must be regarded a critical step preceding \Had injection into 
the \Oad covered (001) plane. Following the resonant field ionisation 
mechanism proposed by Kreuzer and Wang \cite{kreuz}, an oxygen-covered Pt 
surface is expected to be imaged with considerable brightness when using 
molecular NO gas. On the other hand, hydroxyl species formed by reaction of 
\Oad with \Had, are thought to be imaged with less brightness than \Oad, 
mainly because of the lower NO ionisation probability at slightly larger 
critical distances from the  surface.
The above arguments lead to the conclusion that O\Had species are formed
during the NO/H$_2$ reaction. They appear as small clusters with low FIM
brightness in the "defect" region of the (001) layer edge and move
collectively into the flat terrace region. One of the attractive features
of the underlying reaction mechanism is the displacement of hydrogen
islands on top of \Oad rather than exchanging \Oad for O\Had. Alternative
explanations according to which field-adsorbed water clusters are
associated with the occurrence of FIM islands, can be rejected on the basis
of results obtained for the O$_2$/H$_2$ reaction on Pt tips by field ion
appearance potential spectroscopy~\cite{sieben}.

So far the microscopic mechanisms governing the dynamics of the
experimentally observed phenomena described above have not been given
a satisfactory explanation. A possible approach to the microscopic
aspects of these surface reactive phenomena  is provided by 
cellular automaton approaches \cite{boon} or lattice Monte-Carlo modeling 
\cite{zhdanov}. Here we construct a cellular automaton \cite{auto} 
on a two-dimensional square lattice
whose nodes can be occupied by virtual particles with exclusion principle
(no more than one particle per node). The particles are subject to 
probabilistic motion with displacements along the four lattice directions: 
at each time step each particle can hop to any of its four nearest neighboring 
sites. The exclusion principle precludes simultaneous displacements of all 
the particles in the automaton universe in one time step. Therefore a 
sequential updating is implemented: the propagation phase consists of a 
sequence of successive displacements (on the average one per particle) and 
the {\em effective} time step is defined as the sum of $N$ 
automaton time steps, where $N$ is the number of particles in the automaton
universe. Each particle can move to one of its neighboring sites, unless
its destination site is occupied, in which case the particle does not move.
The particles are subject to non-local two-body interactions, which are 
short-range attractive and long-range repulsive, and the amplitude and the 
range of the {\em attractive well} and of the {\em repulsive wall} can be 
tuned parametrically. 

The justification of the automaton for the modeling of the surface phenomena 
observed in the experiments showing adsorbate islands, is as follows:
(i) the processes take place in two-dimensional space, the Pt
surface, and the $(001)$ top plane has square symmetry (bulk-truncate form);
the automaton reproduces these conditions as well as the size of the top 
plane ($30$ sites across);
(ii) the restriction of a single hydroxyl group per site is accounted for by 
the exclusion principle; 
(iii) an hydroxyl group can be identified as an automaton particle; 
(iv) the automaton has intrinsic fluctuations;~(v) the particle dynamics 
depends on the interactions between OH groups
(see caption of Fig.3) which are short-range attractive and long-range 
repulsive~\cite{mikhai}.

In the ``vacuum'' a particle has equal probability to move to any of its four 
neighboring sites; in the presence of other particles, this probability is 
biased according to the occupation of the sites within some {\em neighborhood}
around the particle. The neighborhood is defined as the lattice domain 
surrounding an occupied site and containing all sites within 
interaction range. The bias is such that the sum of the probabilities at each 
site is conserved ($=1$) so that  a positive (negative) interaction along one 
channel induces a repulsive (attractive) interaction along the other channels,
and the propagation direction is drawn according to the probabilities 
evaluated, for the different channels, from the occupation of the lattice 
sites in the neighborhood. 

The updating of the asynchronous automaton at each {\em effective} time 
step consists of a sequential series of operations. 
(i) An occupied site is  chosen randomly; the probability that the particle 
be displaced to one of its four neighboring sites is initially set to $.25$ 
for each channel. (ii) The sites located within the neighborhood of the 
chosen particle are tested for the presence of particles, and the displacement
probabilities are adjusted according to the position and distance of the 
particles within interaction range. (iii) In order to preserve 
the stochastic nature of the procedure, a random number is drawn between
$0$ and $1$ and compared to the cumulated values of the probabilities of
the successive channels; the channel for which the cumulated value exceeds
the random number, is selected. (iv) The particle is displaced by one lattice 
unit along the selected channel, unless the destination site is occupied. 
(v) If so, the particle does not move, and the propagation probability is
transfered to the target particle which memorizes one ``collisional unit''
in the corresponding channel for later updating; the initial probability
for displacement ($.25$) is then biased accordingly.
This step in the algorithm acts as a process with momentum transfer memory.  

The boundary conditions of the automaton universe should be such that they
correspond to the physical situation where hydrogen invades the Pt(001)
surface from the adjacent planes, and particular properties must be assigned 
to the border nodes: a particle located on 
a site within a given distance of a border node is subject to a repulsive 
force which favors the particle displacement along the channels pointing 
inwards the system. This affects any particle residing on a site located at 
a distance equal or smaller than the border interaction range. Circular 
boundary conditions are imposed (i.e. a set of nodes with the boundary

\begin{figure}
\centerline{\psfig{figure=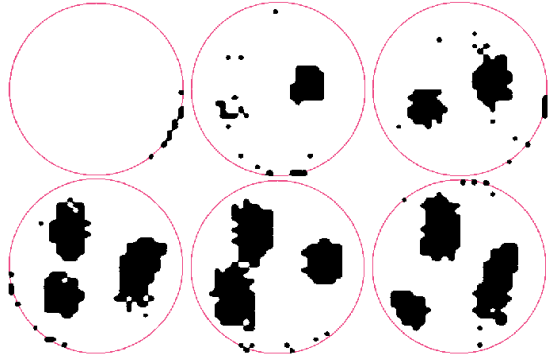,height=5cm}
           }
\vspace{4mm}
\caption{Islands formation and dynamics in automaton with circular open 
boundary (diameter: $30$ nodes). Interaction range and amplitudes: $9$ 
nodes, with attraction on first $2$ nodes and next $3$ nodes,
zero-interaction over next $2$ nodes, and repulsion over last $2$ nodes;
relative values of the amplitudes: $-6:-3:0:+2$ respectively. 
Repulsion from boundary nodes: interaction range $9$ nodes, with relative
amplitude corresponding to $5\%$  bias along the inward direction.
Burst injection rate: $30$ time steps every $100$ time steps.
The time sequence goes from left to right and from top row to bottom row:
time $t=1$ (upper left panel), $600, 1500, 3300, 5100$, and $5700$ (lower
right panel) ($t$ in automaton time step units). An animated version of 
the simulation is available on the web site {\tt http://poseidon.ulb.ac.be}.} 
\label{fig.3}
\end{figure}

\noindent  repulsive effect is chosen to approach a circular 
geometry on the square lattice), and particles are injected in the form of 
bursts over a limited region of the boundary as follows. The circular 
boundary is ascribed a given arbitrary angular reference location where 
from an angle $\alpha$ is chosen randomly. The set of boundary sites in the 
arc spanned between $\alpha$ and a randomly chosen $\Delta \alpha$ define 
the region where particles are injected randomly during a time $\Delta \tau$;
the time intervals between $\Delta \tau$'s are periodically distributed. 
The injection procedure is repeated for the duration of the simulation. 

In Fig.3 we present a sequence of snapshots from the model simulation
starting with a burst injection and leading to the formation of islands
and their subsequent evolution. Comparison with the experimental 
micrographs of Fig.2 shows that the clusters obtained in the automaton 
simulation exhibit the essential features of the observed islands: 
their size ($\sim 30$ sites) and number ($3$ to $4$) have the correct 
order of magnitude, they remain ``autonomous''(when in contact they do 
not merge), and their dynamics reflects the erratic motion of the islands. 
The agreement between the simulation results and the experimental 
observations suggests that the elementary processes on the basis of which 
the automaton is constructed, provide a plausible mechanism for island 
formation and dynamics as observed in FIM experiments whose results were
awaiting satisfactory explanation. The basic mechanism responsible
for the islands formation is governed the balance between attractive and 
repulsive forces between OH$_{ad}$ groups. The relative values of the 
interaction forces given in the caption of Fig.3 give indication as to
what a plausible interaction potential could be between OH$_{ad}$ groups;
furthermore we found the dynamics to be quite sensitive to the value of 
the parameters (for further explorations of surface aggregation phenomena
by automaton simulations, see \cite{bobodha}), and the values used here 
optimize the agreement with the experimental observations. The other crucial 
feature is the role played by fluctuations as intrinsically contained in 
the automaton dynamics. So we conjecture that island formation and
displacements are governed primarily (i) by the balance between lateral 
attractive and repulsive forces between OH$_{ad}$ groups, and (ii) by 
dynamical fluctuations with memory effects caused by frustrated momentum 
transfer.  
 

\begin{center}
{\bf Acknowledgments}
\end{center}

NK thanks the {\em Communaut\'e Fran\c{c}aise de Belgique} (ARC, 
No 96(01-201)) and INTAS (Ukraine 95-0186) for financial support. 
VM gratefully acknowledges a scholarship from ULB.
DH benefited from a grant from the {\em Fonds pour la Formation \`a la 
Recherche dans l'Industrie et l'Agriculture} (FRIA, Belgium). 
JPB acknowledges support by the {\em Fonds National de la Recherche 
Scientifique} (FNRS, Belgium).


\begin{thebibliography}{99}

\bibitem{wintter}
J. Wintterlin, S. V\"{o}lkening, T.V.W. Janssens, T. Zambelli, and G. Ertl, 
{\em Science}, {\bf 278}, 1931 (1997).

\bibitem{zambel}
Z. Zambelli, J.V. Bart, J. Wintterlin, and G. Ertl, 
{\em Nature}, {\bf 390}, 495 (1997).

\bibitem{gorod}
V. Gorodetskii, W. Drachsel, and J. H. Block, 
{\em Ctal. Lett.}, {\bf 19}, 223 (1993).

\bibitem{vosskr}
C. Voss and N. Kruse, 
{\em Appl. Surf. Sci.}, {\bf 87/88}, 127 (1994).

\bibitem{sieben}
B. Sieben, G. Bozdech, N. Ernst, and J. H. Block, 
{\em Surf. Sci.}, {\bf 167}, 352, (1996).

\bibitem{boon}
J.P. Boon, D. Dab, R. Kapral, and A. Lawniczak, 
{\em Phys. Reps.}, {\bf 173}, 55 (1996). 

\bibitem{gauss}
A. Gaussmann and N. Kruse, 
{\em Catal. Lett.}, {\bf 10}, 305 (1991).

\bibitem{vossab}
C. Voss, G. Abend, and N. Kruse (to be published).
 
\bibitem{heinz}
K. Heinz, P. Heilmann, and K. M\"{u}ller, 
{\em Z. Naturf.}, {\bf 32a}, 28 (1977).

\bibitem{vanhove}
M. A. Van Hove, R.J. Koestner, P.C. Stair, J.P. Biberian, L.L. Kesmodel, 
I. Bartos, and G.A. Somorjai, 
{\em Surf. Sci.}, {\bf 103}, 189 (1981); {\em ibid.}, {\bf 103}, 218 (1981).

\bibitem{bonzel}
H.P. Bonzel, G. Broden, and G. Pirug, 
{\em J. Catal.}, {\bf 53}, 96 (1978).

\bibitem{schwa}
K. Schwaha and E. Berthold, 
{\em Surf. Sci.}, {\bf 66}, 383 (1977).

\bibitem{lombar}
S.J. Lombardo, T. Fink, and R. Imbihl, 
{\em J. Chem. Phys.}, {\bf 98},  5526 (1993).

\bibitem{zemly}
D.Y. Zemlyanov, M.Y. Smirnov, V.V. Gorodetskii, and J.H. Block, 
{\em Surf. Sci.}, {\bf 329}, 61 (1995).

\bibitem{kreuz}
H.J. Kreuzer and R.L.C. Wang, 
{\em Z. Phys. Chem.}, {\bf 202}, 127 (1997).

\bibitem{verheij}
L.K. Verheij, M.B. Hugenschmidt, B. Poelsema, and G. Comsa, 
{\em Surf. Sci.}, {\bf 233}, 209 (1990).

\bibitem{zhdanov}
V.P. Zhdanov, {\em Phys. Rev. E}, {\bf 59}, 6292 (1999),
and references therein.

\bibitem{auto}
We use the general terminology {\em cellular automata} for the
model described here by analogy with the {\em lattice gas
automaton} approach used for the microscopic simulation of 
two-dimensional reaction-diffusion systems (ref. \cite{boon} above).
The sequential updating in the present `automaton' makes the model
equivalent to a lattice Monte-Carlo procedure.

\bibitem{mikhai}
A theoretical mesoscopic approach to model traveling nanostrucures in 
surface reactions, based on the assumption of strong attractive 
adsorbate-adsorbate interactions, is presented in 
M. Hildebrand, A.S. Mikhailov, and G. Ertl, 
{\em Phys. Rev. Lett.}, {\bf 81}, 2602 (1998).

\bibitem{bobodha}
J.P. Boon, C. Bodenstein, and D. Hanon, 
{\em Int. J. Mod. Phys. C}, {\bf 9}, 1559 (1998).



\end{thebibliography}
\end{document}